\documentclass[12pt]{article}
\usepackage{epsfig}

\def\beq{\begin{equation}}
\def\eeq#1{\label{#1}\end{equation}}
\def\eeqn{\end{equation}}

\def\beqa{\begin{eqnarray}}
\def\eeqa#1{\label{#1}\end{eqnarray}}
\def\eeqan{\end{eqnarray}}

\let\bar=\overbar

\def\D{{\cal D}}

\def\Dslash{\not{\hbox{\kern-4pt $D$}}}
\def\dslash{\not{\hbox{\kern-2pt $\del$}}}

\def\msb{{\bar{\ssstyle M \kern -1pt S}}}

\def\Title#1{\begin{center} {\Large {\bf #1} } \end{center}}

\AtBeginDocument{

}

\begin{document}

\hfill SLAC-R-860
\vskip 0.5cm

\hfill \hphantom{XXXX}

\Title{Review of Exclusive $B\rightarrow D^{(*,**)}l\nu$ Decays
 -- Branching Fractions, Form-factors and $|V_{cb}|$}

\begin{center}{\large \bf Contribution to the Proceedings of HQL06,\\
Munich, October 16th-20th, 2006}\end{center}

\bigskip\bigskip


\begin{raggedright}  

{\it A. E. Snyder\index{Snyder, A.E.}\\
Stanford Linear Accelerator Center (SLAC)\\
Menlo Park, CA, USA 94041\\
}
\bigskip\bigskip
\end{raggedright}

\section{Introduction}

This paper reviews semileptonic decays of $B$-mesons to states 
containing charm mesons, {\it i.e.}, $D$,\  $D^*$,\ $D^{**}$ and 
possible non-resonant $D^{(*)}n\pi$ states as well. The paper covers 
measurement of branching 
fractions, form-factors and, most importantly, 
the magnitude of the CKM matrix 
element $V_{cb}$. 

I will not attempt a comprehensive review, but will concentrate on reasonably
fresh results and consider mostly exclusive measurements. 
I will also comment on the consistency of the results and what 
needs to be done to resolve the apparent conflicts.

\section{Physics and motivation}

\begin{figure}[htb]
\begin{center}
\epsfig{file=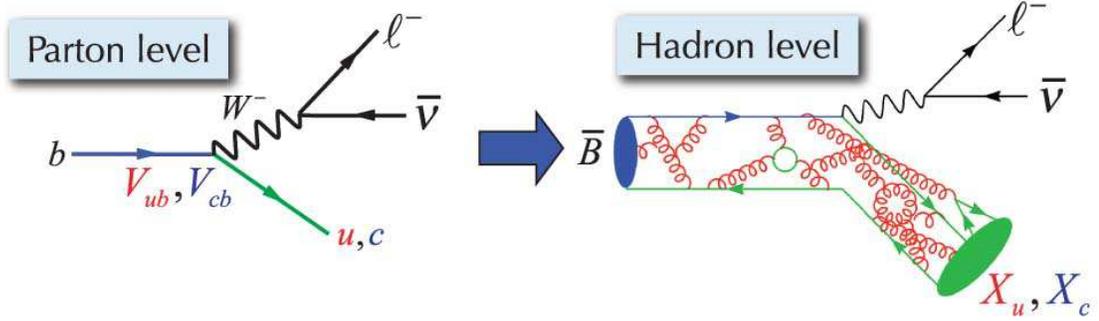,scale=0.9,angle=0}
\caption{Parton level and hadronic level diagrams for semi-leptonic $B$
decay.
}
\label{fig:diagram}
\end{center}
\end{figure}

At the parton level (see Figure~\ref{fig:diagram}) 
the decay rate is simply related to $|V_{cb}|$ by

\begin{equation}
\Gamma=\frac{G_{F}^{2}}{192\pi^{3}}m_b^5|V_{cb}|^{2}
\label{eq:a}
\end{equation}
at tree level. A slightly more complicated formula applies if higher order QCD 
(loop corrections) are considered~\cite{ahmady}. It's still 
thought to be theoretical clean at the parton level, but the parton level process cannot be measured.

Experiment can only measure at the hadron level.
There are two general approaches to 
measuring the decay rate: inclusive measurements in which one sums over the possible
final states $X_c$ and exclusive in which one selects a particular state, 
such as $X_c=D^{*+}$. In the latter case one has to account for the probability (represented by
form-factors) that the 
$c$-quark and the spectator quark combine to form the selected final state 
particle. The form-factors depend on the momentum transfer 
$q^2=(P_l+P_{\nu})^2$ and the formula that relates the
decay rate to $|V_{cb}|$ depends on the final state selected.

\section{$B\rightarrow Dl\nu$}

The decay rate for $B\rightarrow Dl\nu$ is given by
\begin{equation}
\frac{d\Gamma}{dw}=\frac{G_{F}^{2}}{48\pi^{3}}|V_{cb}|^{2}(m_{B}+m_{D})^{2}m_{D}^{3}(w^{2}-1)^{\frac{3}{2}}{\cal F}_{D}^{2}(w)
\label{eq:b}
\end{equation}
where the convention is to use

\begin{equation}
w=\frac{m_{B}^{2}+m_{D}^{2}-q^{2}}{2m_{B}m_{D}}=\frac{E_{D}^{*}}{m_{D}}
\label{eq:c}
\end{equation}
instead of $q^2$ and $\cal F_D$ is the form-factor. Because this
is $0^-\rightarrow 0^-$ transition only one form-factor is needed.

In the heavy quark symmetry (HQS) limit
({\it i.e.} for infinite $c$- and $b$-quark masses) 
${\cal F_D}(w=1)=1$. However, as quarks
are not infinitely heavy, a correction is needed. Lattice QCD calculations
of Hashimoto and collaborators finds ${\cal F_D}(w=1)=1.069\pm 0.026$~\cite{hashimoto}.
Because ${\cal F_D}(1)$ may change as theory matures, it is common practice to
give the experimentally measured quantity ${\cal F_D}(1)\times |V_{cb}|$.

To extrapolate to $w=1$ (or equivalently integrate over the full $w$ range 
with normalization at $w=1$ imposed) the shape of ${\cal F_D}(w)$ is also 
needed. In principle this could be measured, but the rate at $w=1$ vanishes,
so some theoretical input that constrains the form-factor shape has to be deployed.

The most popular (though not the first) 
${\cal F_D} $ parameterization is that of Caprini, Lelloch and Neubert
(CLN)~\cite{clnd} based on Heavy Quark Effective Theory (HQET) and using
dispersion relations to constrain the parameterization. 
The CLN parametrization 
for ${\cal F_D}(w)$ is 
\begin{equation}
{\cal F}_{D}(w)={\cal F_{D}}(1)\times(1-8\rho^{2}z+(51\rho^{2}-10)z^{2}+(252\rho^{2}-84)z^{3}
\label{eq:clnd}\end{equation}
where
$z\equiv(\sqrt{w+1}-\sqrt{2})/(\sqrt{w+1}+\sqrt{2})$
is an improved expansion variable (that converges faster than $w-1$).

Some other parameterizations are provide by Boyd, Grinstein 
and Lebid (BGL)~\cite{bgl} and LeyYaonac, Oliver 
and Raynal (LeYor)~\cite{leyor}. 

In all these parameterizations we are left with only one parameter to fit
-- $\rho^2$ -- which is the derivative of ${\cal F_D}(w)$  {\it w.r.t.}
$w$ at $w=1$.

\begin{figure}[htb]
\begin{center}
\epsfig{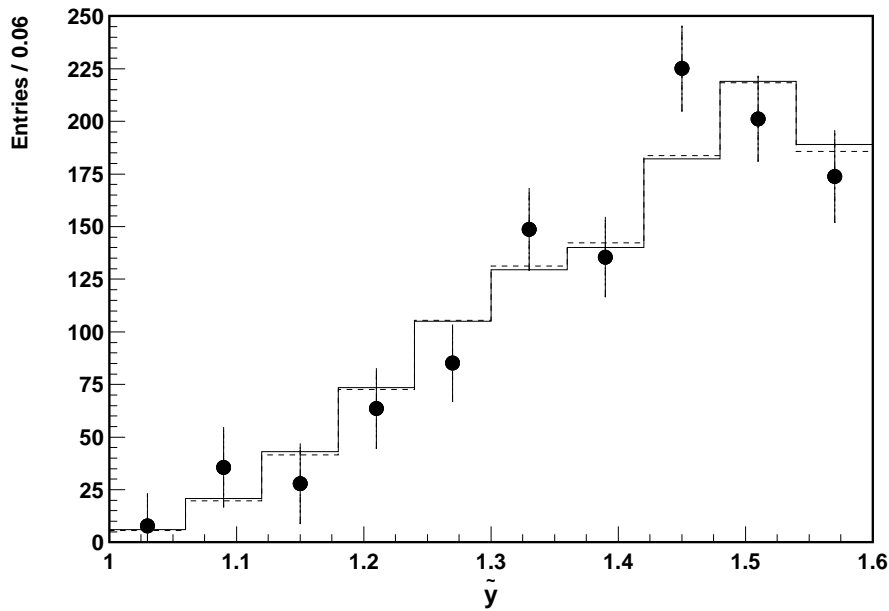}
\caption{Background subtracted $w$-distribution (called ${\hat y}$ here) as 
measured by BELLE. Points
are data and histograms are fits.
}
\label{fig:belled}
\end{center}
\end{figure}

The $D\rightarrow Dl\nu$ analysis has been carried out by the CLEO~\cite{cleod} 
and BELLE~\cite{belled} collaborations. CLEO used both 
${\bar B}^-\rightarrow D^0l^-{\bar\nu}$ and $B^-\rightarrow D^+l^-{\bar\nu}$
modes, while BELLE only used the $B^-\rightarrow D^+l^-\bar\nu$. 
Figure~\ref{fig:belled} shows the background subtracted $w$-distribution 
obtained by BELLE (note in BELLE's notation $w$ is called ${\hat y}$). The signal
is extracted using the discriminating variable $cos\theta_{BY}$ defined and described
in detail in section~\ref{sec:dsln}.

They fit by minimizing $\chi^2$ given by
 
\begin{equation}
\chi^{2}=\sum_{i}\left(\frac{N_{i}^{obs}-\sum_{j}\varepsilon_{ij}N_{j}(N,\rho^{2})}{\sigma_{i}}\right)^{2}
\label{eq:bellechisq}
\end{equation}
where $N_j$ is the number predicted based on Eq.(\ref{eq:b}) and 
$\varepsilon_{ij}$ is the efficiency for an event truly in bin $i$ to be
detected in bin $j$. The errors $\sigma_i$ includes both the uncertainty in the
observation and the uncertainty in the efficiency matrix $\varepsilon_{ij}$

The fit parameters are the number of events $N$ and
the slope parameter $\rho^2$. The fit using the CLN parameterization
(Eq.(\ref{eq:clnd})) is shown
as solid histogram in the figure. The dashed histogram represents the result
when a simple linear parameterization is used instead of CLN. The two fits are
not distinguishable.

CLEO does something similar. The results of both experiments is given table
~\ref{tab:dresults} along with the HFAG averages\footnote{HFAG averages in
includes older result from ALEPH} as of summer 2006.

\begin{table}
\begin{center}\begin{tabular}{|c|c|c|c|}
 \hline 
 { Exp}&
 { ${ \cal B}(B^{-}\rightarrow D^{0}l^{-}\bar{\nu})(\%)$}&
 { $\rho^{2}$(CLN)}&
{ ${\cal F}(1)\times|V_{cb}|(10^{-3})$}\tabularnewline
\hline
\hline 
{ CLEO}&
{ $2.21\pm0.13\pm0.19$}&
{ $1.27\pm0.25$}&
{ $44.8\pm5.8\pm3.$}\tabularnewline
\hline 
{{ BELLE}}&
{ $2.13\pm0.12\pm0.39$}&
{ $1.12\pm0.22$}&
{ $41.1\pm4.4\pm5.1$}\tabularnewline
\hline
\hline 
HFAG&
$2.12\pm 0.20$&
$1.17\pm 0.18$&
$42.4\pm 4.5$\tabularnewline
\hline 
\end{tabular}
\caption{CLEO and BELLE results for branching fraction, $\rho^2$ and
$|V_{cb}|$ and HFAG averages}
\label{tab:dresults}
\end{center}
\end{table}

The BELLE and CLEO results are consistent. The uncertainties are large and BELLE assigns a more 
conservative systematic than CLEO.

\section {$B\rightarrow D^*l\nu$}
\label{sec:dsln}

\begin{figure}[htb]
\begin{center}
\epsfig{file=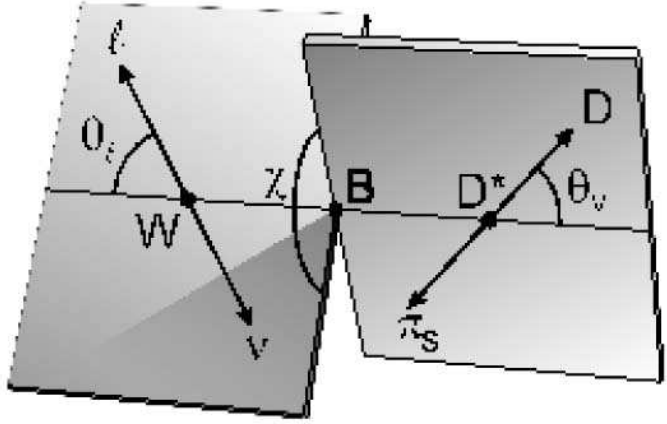,scale=1.0,angle=0}
\caption{Definition of the angles 
$\theta_l,\ \theta_V$ and $\chi$ that describe the
decay $B\rightarrow D^*l\nu$.
}
\label{fig:dslnuang}
\end{center}
\end{figure}

The analysis of $B\rightarrow D^*l\nu$ is more complex than $Dl\nu$. There are three form-factors 
called $A_1,\ A_2$ and $V$.
To separate them  an analysis in the three angles ($\theta_l,\ \theta_V$ and $\chi$) and $w$ is needed. 
 The angles are defined in Figure~\ref{fig:dslnuang}. 
 
The decay rate in terms of the four kinematic variables $w,\ \theta_l,\ \theta_V$ and $\chi$ is given by

\begin{equation}
\frac{d\Gamma}{dw\  dcos_{l}\  dcos_{V}\  d\chi}=K|V_{cb}|^{2}q^{2}p_{D^{*}} 
\times 
\label{eq:dslna}
\end{equation}
\begin{equation}\nonumber
  \{ H_{+}^{2}(1-cos\theta_{l})^{2}sin^{2}\theta_{V}+H_{-}^{2}(1+cos\theta_{l})^{2}sin^{2}\theta_{V} +4H_{0}^{2}sin^{2}\theta_{l}cos^{2}\theta_{V}\
\end{equation}
\begin{equation} 
-2H_{+}H_{-}sin^{2}\theta_{l}sin^{2}\theta_{V}cos2\chi -4H_{0}(H_{+}\left(1-cos\theta_{l})-H_{-}(1+cos\theta_{l})\right)sin\theta_{V}cos\theta_{V}cos\chi\} 
\end{equation}
where $H_\pm$ and $H_0$ are helicity amplitudes related to the form-factors by
\begin{equation}
H_{\pm}=-(m_{B}+m_{D^{*}})A_{1}(w)\pm\frac{2p_{D^{*}}m_{B}}{m_{B}+m_{D^{*}}}V(w)
\end{equation}
\begin{equation}
H_{0}=-\frac{m_{B}+m_{D^{*}}}{m_{D^{*}}\sqrt{q^{2}}}(m_{D^{*}}(wm_{B}-m_{D^{*}})A_{1}(w)-\frac{4m_{B}^{2}p_{D^{*}}^{2}}{(m_{B}+m_{D^{*}})^{2}})A_{2}(w).
\end{equation}
In principle, by doing a spin-parity analysis in bins of $w$ the form-factors could be measured without any model dependence. 
In practice the statistics are inadequate and we have to resort to some parameterization.

HQET relates the three form-factors describing the decay to a single common form-factor called the Isgur-Wise function. The 
relationships are
\begin{equation}
A_{2}(w)=\frac{R_{2}(w)}{R_{*}^{2}}\frac{2}{w+1}A_{1}(w)
\end{equation}
\begin{equation}
V(w)=\frac{R_{1}(w)}{R_{*}^{2}}\frac{2}{w+1}A_{1}(w)
\end{equation}
\begin{equation}
A_{1}(w)=R_{*}\frac{w+1}{2}h_{A_{1}}(w)\rightarrow R_{*}\frac{w+1}{2}\xi(w)
\end{equation}
where $R_{*}=2\sqrt{m_{B}m_{D*}}/(m_{B}+m_{D^{*}}). $

The CLN~\cite{clnd} formalism is again the most popular. 
They provide the $w$ dependence of $R_1(w)$ and $R_2(w)$ and a parameterization
in terms of the slope $\rho^2$ at $w=1$ of the common form-factor $h_{A_1}(w)$. In fitting 
for the form-factors the intercepts $R_1(w=1),\ R_2(w=1)$ and $\rho^2$ are taken as the independent parameters.

The CLN parameterizations are

\begin{equation}
R_{1}(w)=1.27-0.12(w-1)+0.05(w-1)^{2},
\label{eq:clnr1}
\end{equation}
\begin{equation}
R_{2}(w)=0.79+0.15(w-1)-0.04(w-1)^{2}
\label{eq:clnr2}
\end{equation}
for the form-factor ratio parameters and
\begin{equation}
h_{A_{1}}(w)=h_{A_{1}}(1)(1-8\rho^{2}z+(53\rho^{2}-15)z^{2}-(231\rho^{2}-91)z^{3}
\label{eq:clndstff}
\end{equation}
for the common `Isgure-Wise' like form-factor.

Outside the heavy quark limit where $h_{A_1}(w)=\xi(w)$ and the value
at $w=1$ is 1.0, $h_{A_1}(w=1)$ has to be taken from theory. The best estimate comes 
from lattice QCD~\cite{hashimoto}. It is $0.919^{+0.030}_{-0.035}$. 

\begin{figure}[htb]
\begin{center}
\epsfig{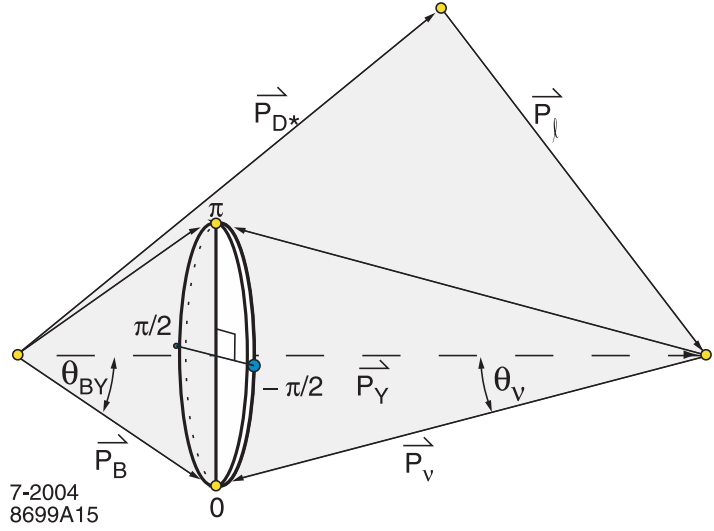}
\caption{
Kinematics of $B\rightarrow D^*l\nu$ that allow $cos\theta_{BY}$ to be reconstructed.
}
\label{fig:reco}
\end{center}
\end{figure}

BABAR measures the decay rates as function of $\theta_l,\ \theta_V,\ \chi$ and $w$. 
Because of the missing neutrino these cannot be
directly obtained from the measured tracks. Thus, a partial reconstruction technique, 
(illustrated in Figure~\ref{fig:reco}), 
which allows a reasonable accurate approximation to be made, is used. 
Using the kinematic constraints expressed in Figure~\ref{fig:reco} the cosine
of the angle between the $B$ direction and the direction of the lepton-$D^*$ system ($\mathbf{P}_{Y}=\mathbf{P}_{D*}+\mathbf{P}_{l}  $) can be obtained as follows:
\begin{equation}
cos\theta_{BY}=\frac{2E_{B}E_{D^{*}+l}-m_{B}^{2}-m_{D^*+l}^{2}}{2|\mathbf{P}_{D^{*}}|\mathbf{P}_{Y}|}
\end{equation}
where lepton and $D^*$ momenta are measured and $|\mathbf{P}_B|$ can be estimated from the beam energies.
Note the same construction is used for $Dl\nu$ with, of course, 
$D\rightarrow D^*$ 

The azimuthal angle $\phi_{BY}$ of $\mathbf{P}_B$ around the $Y$ direction is undetermined.
The kinematic variables 
$cos\theta_l,\ cos\theta_V,\ \chi$ and $w$ can be calculated for any choice of $\phi_{BY}$.
Averaging over $\phi_{BY}$ gives reasonable estimators of their values.
BABAR uses points at $\phi_{BY}=0,\ \pi,\ \pm \pi/2$ weighted by the $B$ production angular distribution
($\propto sin^2\theta_B$) to perform this average.

Only and CLEO~\cite{CLEO} and BABAR~\cite{BABARA,BABARB} have attempted form-factor measurements. BABAR uses two method: 
a likelihood fit to the full $4-d$
distribution of the kinematics variables ($w,\ \theta_l,\ \theta_V,\ \chi$) and a simultaneous fit to the 
one-dimensional $w$, $cos\theta_l$
and $cos\theta_V$ projections. CLEO also does a 4-d likelihood fit. I'll describe the BABAR methods in some detail.

For the $4-d$ likelihood fit it's difficult to construct the full, correlated PDF 
(including efficiency and resolution) for the measured variables 
$w,\ \theta_l,\ \theta_V$ and $\chi$, so BABAR resorts to the `integral method' that avoids the need to know this 
complicated PDF. With the integral method only the integral of the efficiency and the theoretical 
PDF Eq.(\ref{eq:dslna}) are needed.

The extended likelihood (include resolution) is given by
\begin{equation}
log{\cal L}=\sum_e log\tilde{F}(\tilde{\Omega}_{e}|\mu)-\int d\tilde{\Omega}\tilde{F}(\tilde{\Omega}|\mu)
\label{eq:loglike}
\end{equation}
where $\tilde{\Omega}$ represents measured quantities and 
$\tilde{F}(\tilde{\Omega}|\mu)$ represents their PDF for parameters $\mu$.
The sum is over events. Using the approximation
\begin{equation}
\tilde{F}(\tilde{\Omega}|\mu)\approx F(\tilde{\Omega}|\mu)\times\frac{\tilde{F}(\tilde{\Omega}|\mu_{mc})}{F(\tilde{\Omega}|\mu_{mc})}
\end{equation}
we obtain
\begin{equation}
log{\cal L}\approx\sum logF(\tilde{\Omega}_{e}|\mu)-\tilde{I}(\mu,\mu_{mc})
\label{eq:apxlike}
\end{equation}
with
\begin{equation}
\tilde{I}(\mu,\mu_{mc})=\int{d\tilde{\Omega}F(\tilde{\Omega},\mu)\times
\left(\frac{\tilde{F}(\tilde{\Omega}|\mu_{mc})}{F(\tilde{\Omega}|\mu_{mc})}\right)}
\end{equation}
which can be obtained by Monte Carlo integration as
\begin{equation}
\tilde{I}(\mu,\mu_{mc})\approx\frac{1}{N_{mc}}\sum\frac{F(\tilde{\Omega}_{imc}|\mu)}{F(\tilde{\Omega}_{imc}|\mu_{mc})}.
\end{equation}
It can be shown that for $\mu_{true}$ equal $\mu_{mc}$, the result of this approximation is unbiased. 
The procedure can be iterated by re-weighting to get $\mu_{mc}\approx\mu_{fit}$ which is close enough. Extra contributions to
the error from MC statistics need to be evaluated, 
but knowledge of $4-d$ efficiency-resolution 
function is not required. 
Background is subtracted event-by-event 
using MC events to avoid breaking the factorization that leads to
Eq.(\ref{eq:apxlike}).

\begin{figure}
\begin{center}
\epsfig{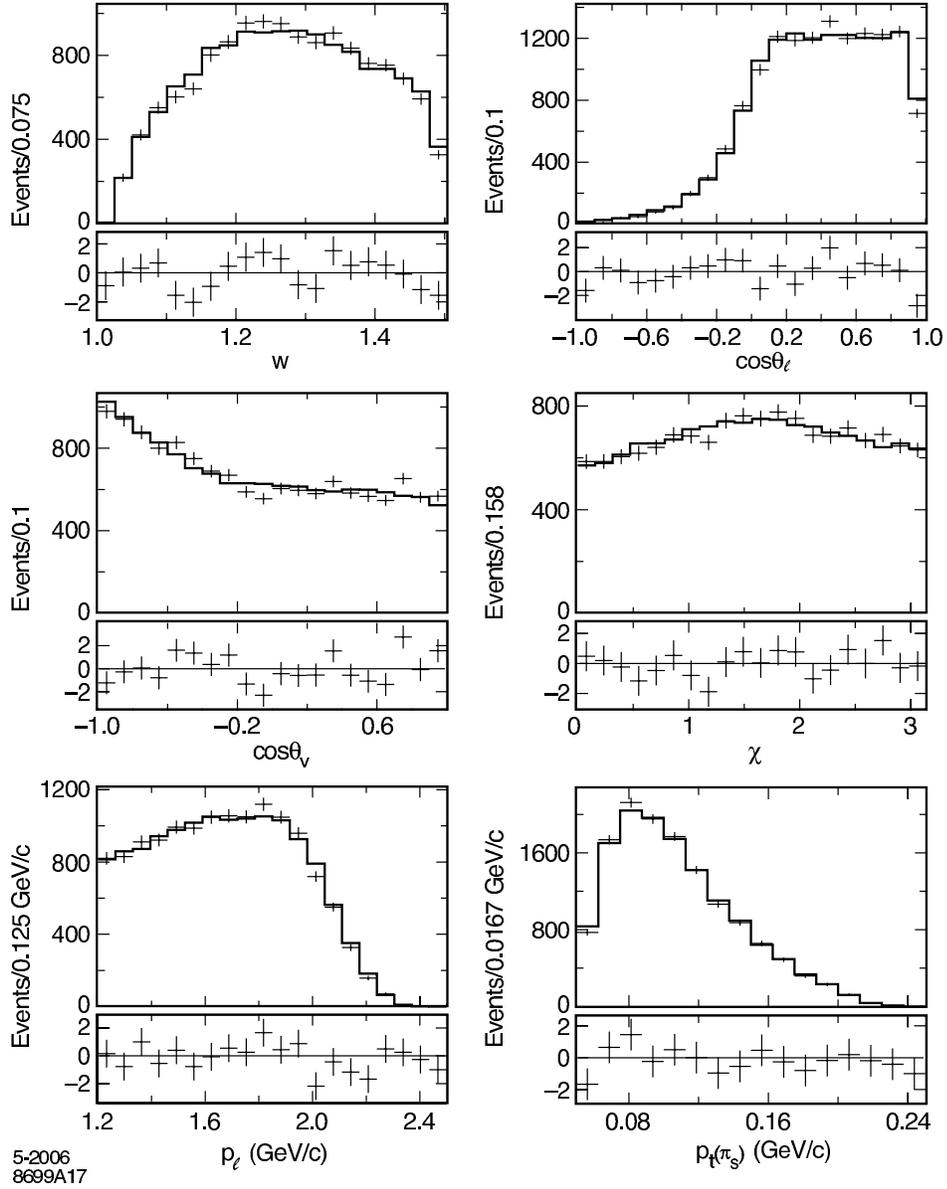}
\caption{Distribution of $w,\ cos\theta_l,\ cos\theta_V,\ \chi$ as well
as lepton momentum $p_l$ and ``slow'' pion momentum $p_t(\pi_s)$.
}
\label{fig:uncut}
\end{center}
\end{figure}

\begin{figure}
\begin{center}
\epsfig{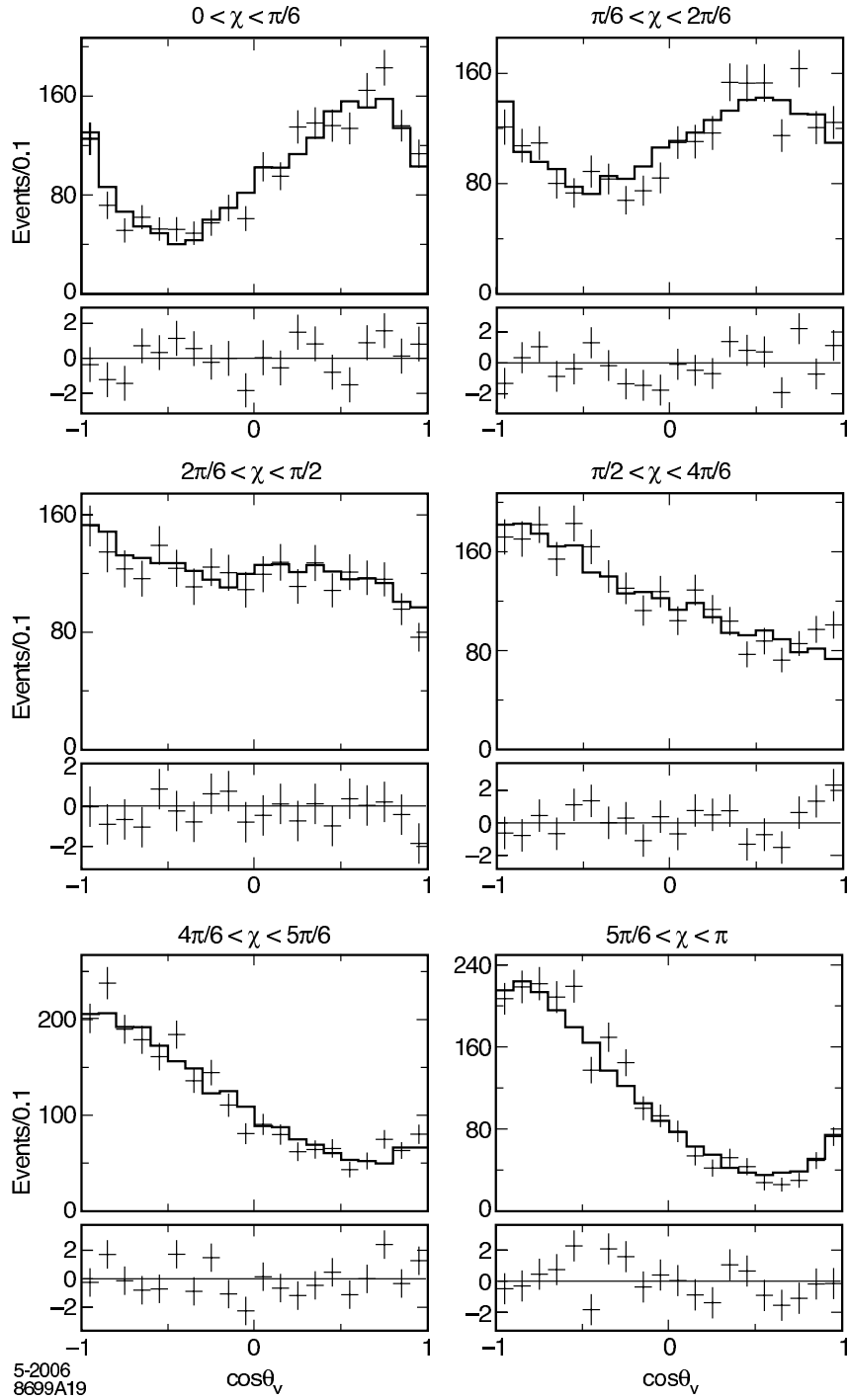}
\caption{Distribution of $\chi$ for six cuts on $cos\theta_V$.
}
\label{fig:cut}
\end{center}
\end{figure}

Since there is no explicit PDF $\tilde{F}$ constructed, the MC is re-weighted to fitted values of $R_{1}(1),\  R_{2}(1)$ 
and $\rho^{2}$ (histogram) and compared to the data (points) to see if fit is good.
Figure~\ref{fig:uncut} shows the four projects and Figure~\ref{fig:cut} shows the $\chi$ projection for six bins
of $cos\theta_V$. The agreement with the fit represented by re-weight MC is excellent and reproduces even the details of 
the interference in the $\chi$ vs. $cos\theta_V$ plots.


The results\footnote{These are for so called baseline case where $R_1(w)=R_1(1)$ and $R_2(w)=R_2(1)$  are
taken to be constant. The results using the CLN or other predictions for the $w$
dependence of $R_1$ and $R_2$ are not much different.} are

\begin{equation}
R_{1}(w)=1.396\pm 0.046\pm 0.027
\end{equation}
\begin{equation}
R_{2}(w)=0.885\pm 0.046\pm0.013
\end{equation}
\begin{equation}
\rho^{2}=1.145\pm 0.066\pm0.035
\end{equation}

The BABAR results are consistent with the pioneering CLEO measurement 
of $R_{1}=1.18\pm0.30\pm0.12,\ R_{2}=0.71\pm0.22\pm0.07 $.
The slope parameter $\rho^2$ also agrees when equivalent parameterizations
of $h_{A_{1}}(w)$ are used. The values are also consistent with theoretical 
expectations. Since, $|V_{cb}|$ is highly sensitive to $R_1$ and $R_2$, this measurement leads
to substantial reduction in the error achievable on it.

The second BABAR method works with projections in $w,\ cos\theta_l$ and $cos\theta_V$. The dihedral angle $\chi$ is not 
used because it has little sensitivity when one integrates over the other angles. The projection method
is not as statistical powerful for the form-factors as the full $4-d$ fit, however, it has the advantage that
the background can be estimated and removed in a manner that is nearly independent of the MC. This allows higher multiplicity
$D$-decays to be used. The $4-d$ method only used the $D\rightarrow K\pi$ in order to keep the systematic error from
dependence on MC simulation of the background shape under control.

The projection method divides each variable into ten bins and fits the $cos\theta_{BY}$ distribution in each bin 
to obtain estimates of the signal and background {\it in that bin}. The shape of the $cos\theta_{BY}$ distribution for
background and signal is taken from MC and whenever possible from 
data control samples, but no assumption about the shape of the
background distributions in $w,\ cos\theta_l$ or $cos\theta_V$ is used. 
Thus the resulting projection plots are only
weakly dependent on the MC simulation.

\begin{figure}
\begin{center}
\epsfig{file=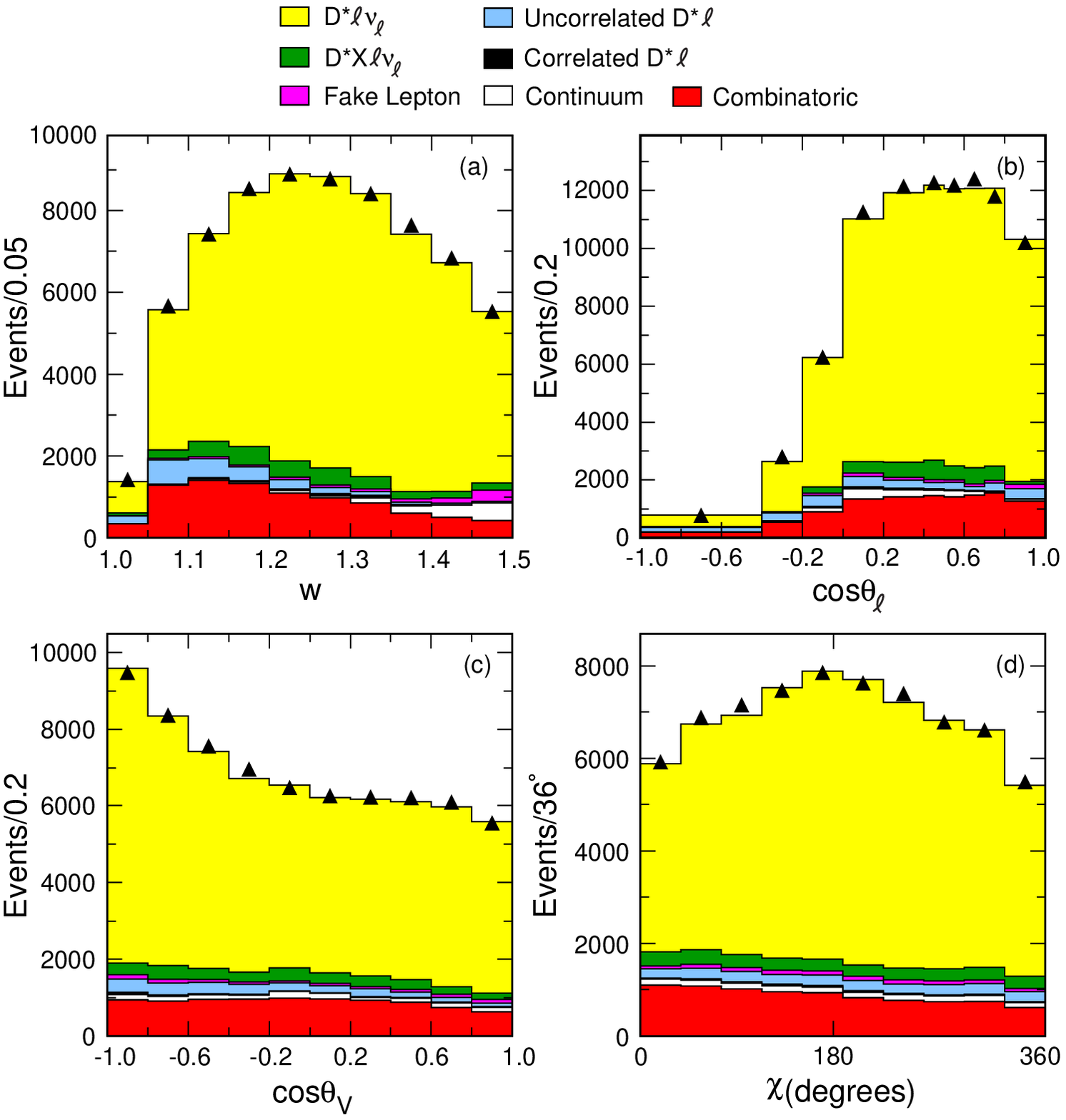,scale=1.0,angle=0}
\caption{
Projection results for $w,\ cos\theta_l,\ cos\theta_V$. The angle $\chi$ was not part of the fit.
}
\label{fig:proj}
\end{center}
\end{figure}

Figure~\ref{fig:proj} shows the results of simultaneous
fits to the projection plots. The projection distributions are correlated since they share
the same events; this correlation is taken into account in the fitting procedure.
The triangles represent that data, the yellow the
signal and the other colors the results of the $\cos\theta_{BY}$ fits for the background. The $\chi$ projection is
not fit, but is only shown for completeness. The fits look good.

For BABAR's final result they combine the results of the two method. 
The results are

\begin{equation}
  {\cal F}(1)|V_{cb}=(34.68\pm 0.32\pm 1.15)\times 10^{-3} 
\end{equation}
\begin{equation}
  \rho^2=1.179\pm 0.048 \pm 0.028
\end{equation}
\begin{equation}
  R_1=1.417\pm 0.061\pm 0.044
\end{equation}
\begin{equation}
  R_2=0.836\pm 0.037\pm 0.022
\end{equation}
where errors are statistical and systematic respectively. In this case
the CLN form (Eqs.~\ref{eq:clnr1}-\ref{eq:clnr2}) for $R_1(w)$ and $R_2(w)$ has been used.

This currently yields the best
exclusive measurement of $|V_{cb}|=(37.74\pm 0.35\pm 1.24^{+1.32}_{-1.44})\times 10^{-3}$. The additional error is 
from the theoretical predictions of ${\cal F}(1)$.

\begin{figure}
\begin{center}
\epsfig{file=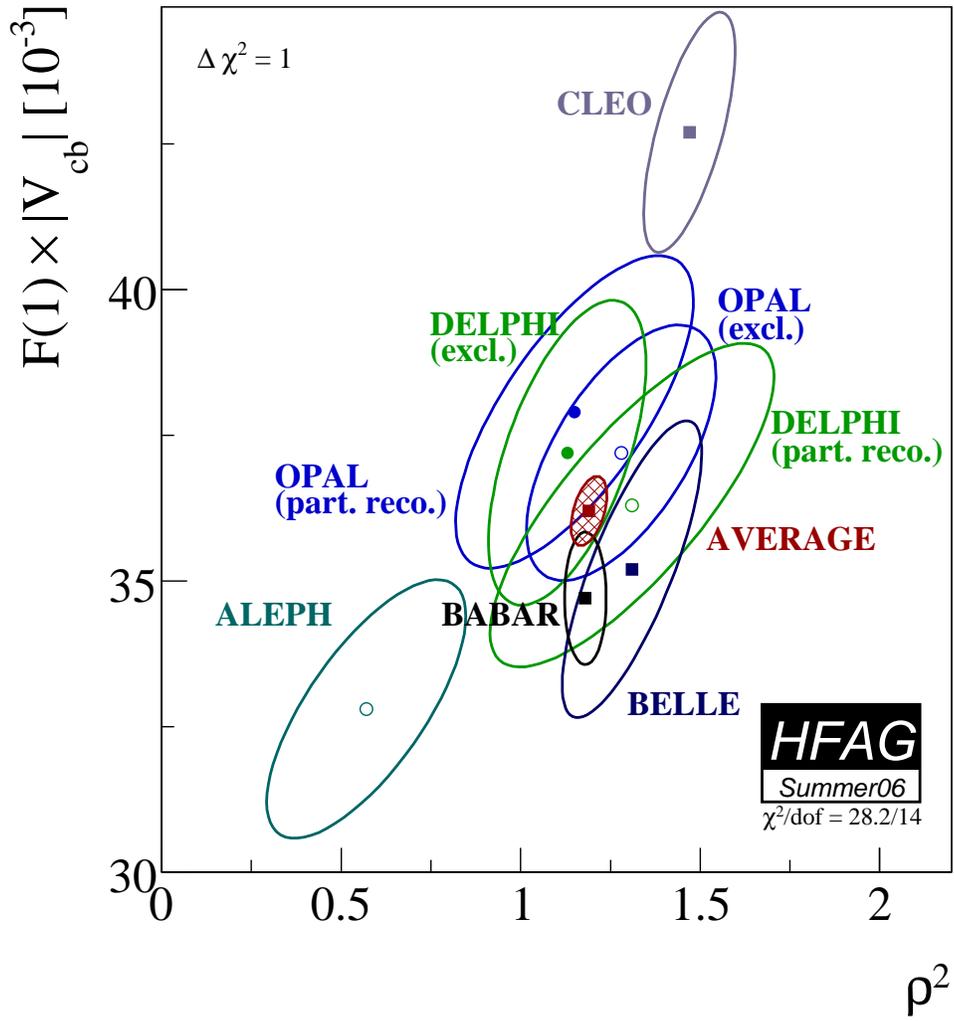,scale=0.7,angle=0}
\caption{
Summer `06 plot of ${\cal F}(1)|V_{cb}|$ vs. $\rho^2$ from HFAG.
}
\label{fig:hfag}
\end{center}
\end{figure}

\begin{figure}
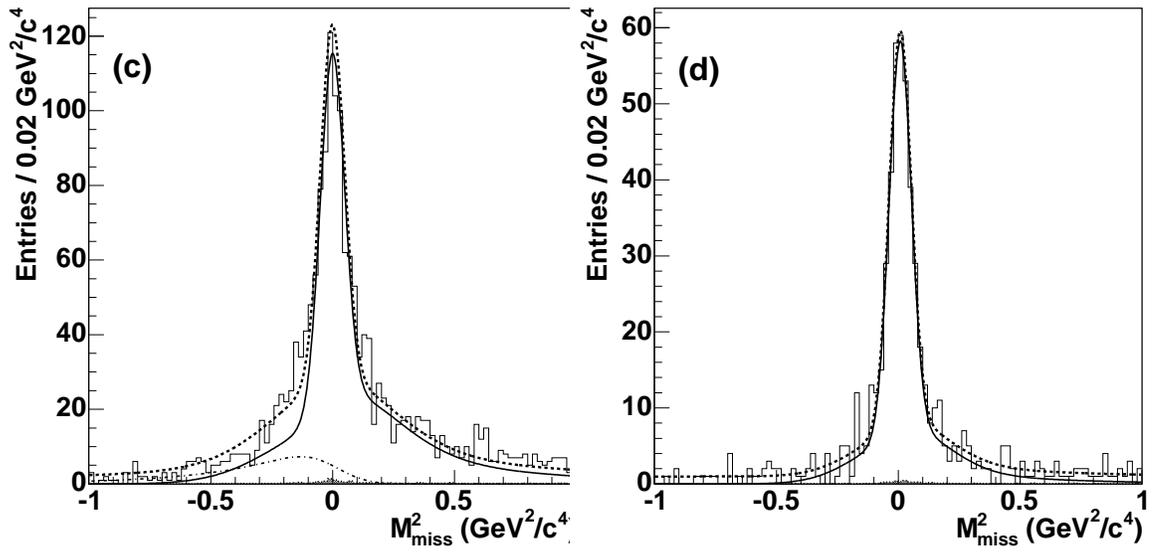

\begin{center}
\epsfig{file=figs/fig2_bc_dstlnu.epsi,scale=0.4,angle=0}\epsfig{file=figs/fig2_bn_dstlnu.epsi,scale=0.4,angle=0}
\caption{
Missing mass against $D^*l\nu$ in $B^{\pm}$ recoil sample (left) and in $B^0$ sample (right).
}
\label{fig:belledslnrec}
\end{center}
\end{figure}

HFAG has averaged the $|V_{cb}|-\rho^2$ results from six experiments. The summer 2006 average, which includes
the BABAR improvement of $R_1$ and $R_2$ to the results of other experiments, is shown in Figure~\ref{fig:hfag}.
The poor $\chi^2$ ($28/14$) is mostly due to the CLEO and ALEPH measurements. While BABAR is the single best measurement,
the others make a significant contribution to the world average.

There is a persistent conflict between the $D^*l\nu$ branching fraction measured with charged and neutral $B$-mesons.
In PDG 2006 average is $5.34\pm 0.20\%$ for neutrals and  $6.5\pm 0.5\%$ for charged $B$'s. This difference would
represent isospin violation, which is $\it a priori$ very unlikely. At this point it's only $\sim 2\sigma$, so we
can hope the difference will ``regress-to-the-mean'' as new measurements come in.

BELLE~\cite{BELLEA} has used a ``recoil'' technique to measure both $D^{*0}l^+\nu$ and $D^{*0}l\nu$ recoiling against a 
fully reconstructed $B^-$ or $\bar{B}^0$. 
While the sample is quite small, this method substantially reduces the background and thus potentially
the systematic errors. BELLE has not yet fully exploited this potential as they do not give a systematic error on these
branching fractions. 

Figure~\ref{fig:belledslnrec} shows BELLE's missing mass plots. The branching fractions they 
obtain are
\begin{equation}
B(\bar{B}^{0}\rightarrow D^{*}l\nu)=4.7\pm0.24\%,\ B(B^{+}\rightarrow D^{*}l\nu)=6.06\pm0.25\% 
\label{eq:beldslnbf}
\end{equation}
which is $\sim 4\sigma$ discrepancy,
...but of course  only the statistical error is taken into account. 
This method has the potential to shed considerable light on the issue, if the systematic uncertainties
can be understood.

\section{Higher mass states: $D^{**},\ D^{(*)}\pi$, etc.}

\begin{figure}
\begin{center}
\epsfig{file=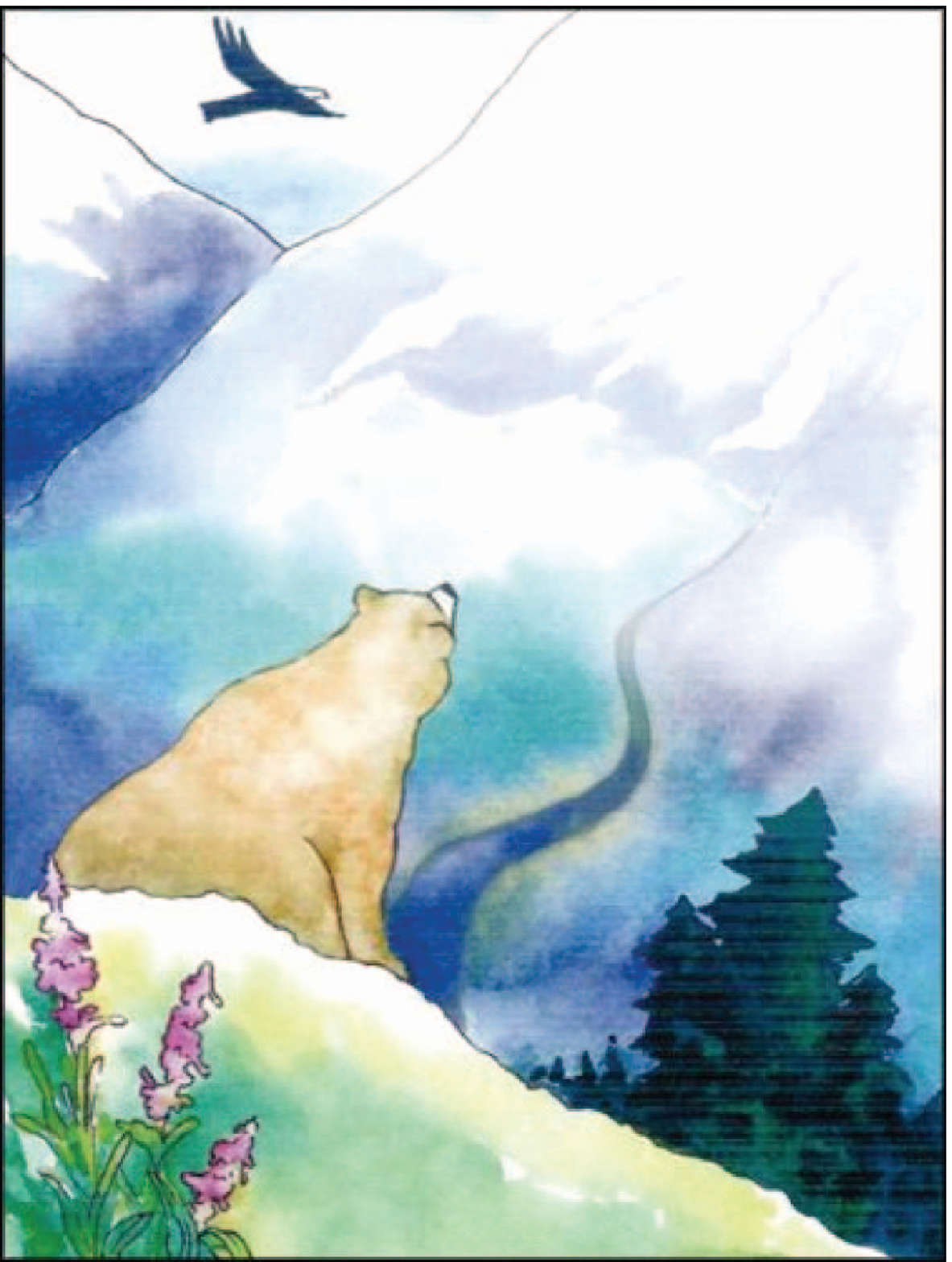,scale=0.5,angle=0}\epsfig{file=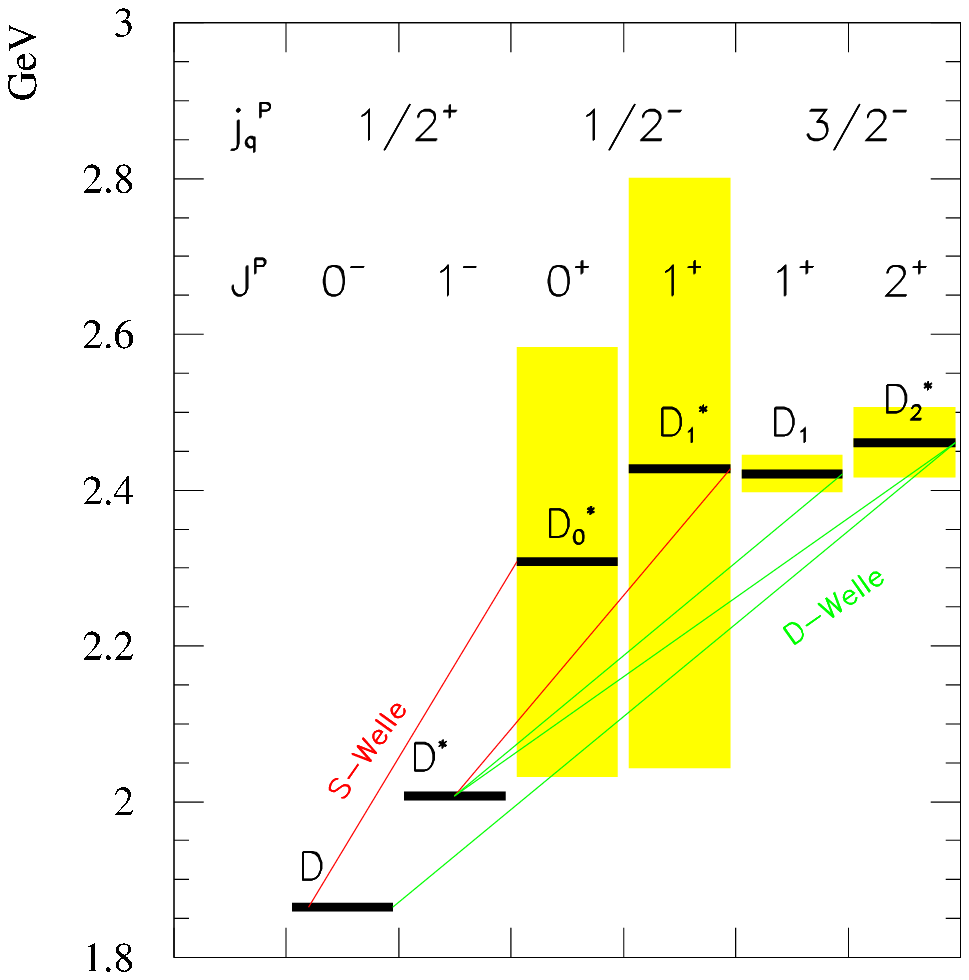,scale=0.9,angle=0}
\caption{
``...to see what he could see...and high mass charm states we might see... ~\cite{bear}
}
\label{fig:bear}
\end{center}
\end{figure}

\begin{figure}
\begin{center}
\epsfig{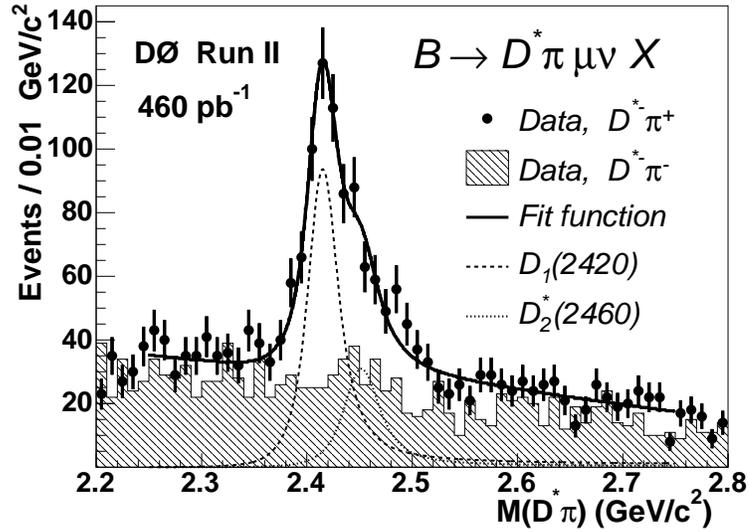}
\caption{
$D^*\pi$ mass distribution from $D^0$.
}
\label{fig:D0}
\end{center}
\end{figure}

One might ask {\it w.r.t}  the semi-leptonic decays to the higher mass charm states ``why bother?'' 
It seems unlikely anything very exciting will be found.

One answer might be simply ``because they're there.'' In general, it's good idea to try to understand the decay modes
of particles like the $B$ as completely as possible...and you'll never know what might be lurking there,
if you don't look...we need to see what we can see (see Figure~\ref{fig:bear}).

A more practical answer is ``engineering.'' The so-called $\D^{**}$ states 
and $D^{*}n\pi$ are the dominant backgrounds to $Dl\nu$ and
$D^*l\nu$ and the lack of understanding of these backgrounds is a major source of systematic error for $V_{cb},\ \rho^2$
and the form-factor ratios $R_1$ and $R_2$.

HQET also makes predictions about the form-factors of excited $D$-mesons and perhaps lattice QCD can too. Measurements
in the high mass sector can test the theory that underlies the extraction of $|V_{cb}|$ and are of some interest in
themselves.

What could be there? In terms of resonance, this also shown in Figure~\ref{fig:bear} where
the predictions for the resonant structure based on HQET are given. There are two narrow resonances -- the $D_1(2420)$ and 
the $D_2^*(2460)$ which should be relatively easy to detect and, in fact, are already seen in semi-leptonic $B$-decay. 
The wide $D_0^*$ and $D_1^*$ will
be hard to distinguish from the non-resonant $D^{(*,**)}n\pi$ contributions.

The D0 experiment~\cite{D0} has observed the narrow states in $B\rightarrow D^*\pi l\nu+X$. The measurement does not strictly
correspond to exclusive $D^*\pi$ though that's likely to be a major component of the signal. The $D^*\pi$ mass plot
is displayed in Figure~\ref{fig:D0}. The $D_2(2460)$ corresponds to the shoulder on the right side of the bump. Only 
the product production$\times$branching fraction are directly measured, they are
\begin{equation}
B\rightarrow \bar{D}_1l\nu,\ \bar{D}_{1}^{0}\rightarrow D^{*-}\pi^{+}=0.087\pm0.007\pm0.014,
\end{equation}
\begin{equation}
B\rightarrow\bar{D}_2l\nu,\ \bar{D}_{2}^{*0}\rightarrow D^{*-}\pi^{+}=0.035\pm0.007\pm0.008,
\end{equation}
\begin{equation}
ratio\ \bar{D}_{2}^{*0}/\bar{D}_{1}^{0}=0.39\pm0.09\pm0.12.
\end{equation}
Using $b\rightarrow B^-=0.39\pm 0.09\pm 0.12$ and isospin symmetry, the absolute branching fractions can be
estimated as $0.33\pm 0.06\%$ for the $D_1$ and $0.44\pm 0.16\%$ for the $D_2^*$.

The OPAL~\cite{OPALdsdst} collaboration does a similar measurement, but is unable to see the $D_2$. They get
\begin{equation}
{\cal B}(b\rightarrow\bar{B})\times{\cal B}(D_{1}^{0}l^{-}\bar{\nu})
\times{\cal B}(D^{*+}\pi^{-})=(2.64\pm0.79\pm0.39)\times10^{-3}.
\end{equation}
For the $D_2$ they get
$(0.26\pm0.59\pm0.35)\times10^{-3}$ which is consistent with zero.
\begin{figure}
\begin{center}
\epsfig{file=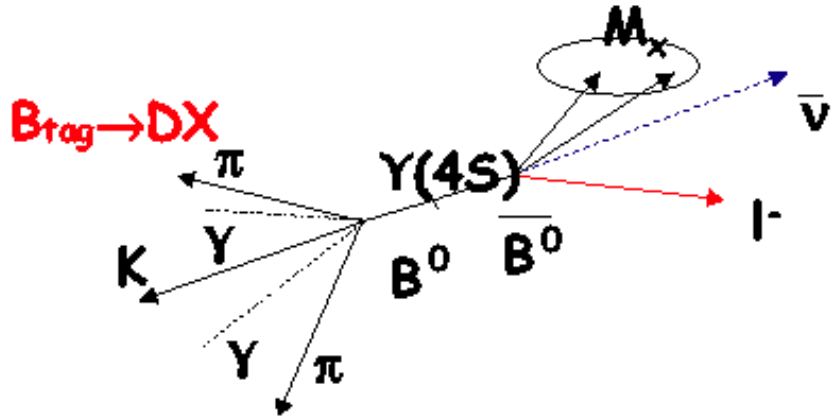,scale=0.8,angle=270}
\caption{
Illustration of reconstruction missing mass $M_{miss}$ in the system recoiling against a fully reconstructed $B$ and
some hadrons (represented by $M_X$). 
For the exclusive channels of interest $M_X$
is composed of $D^{(*)}\pi$ and $M_{miss}=M_{\nu}=0$ for signal.
}

\label{fig:recoil}
\end{center}
\end{figure}

\begin{figure}
\begin{center}
\epsfig{file=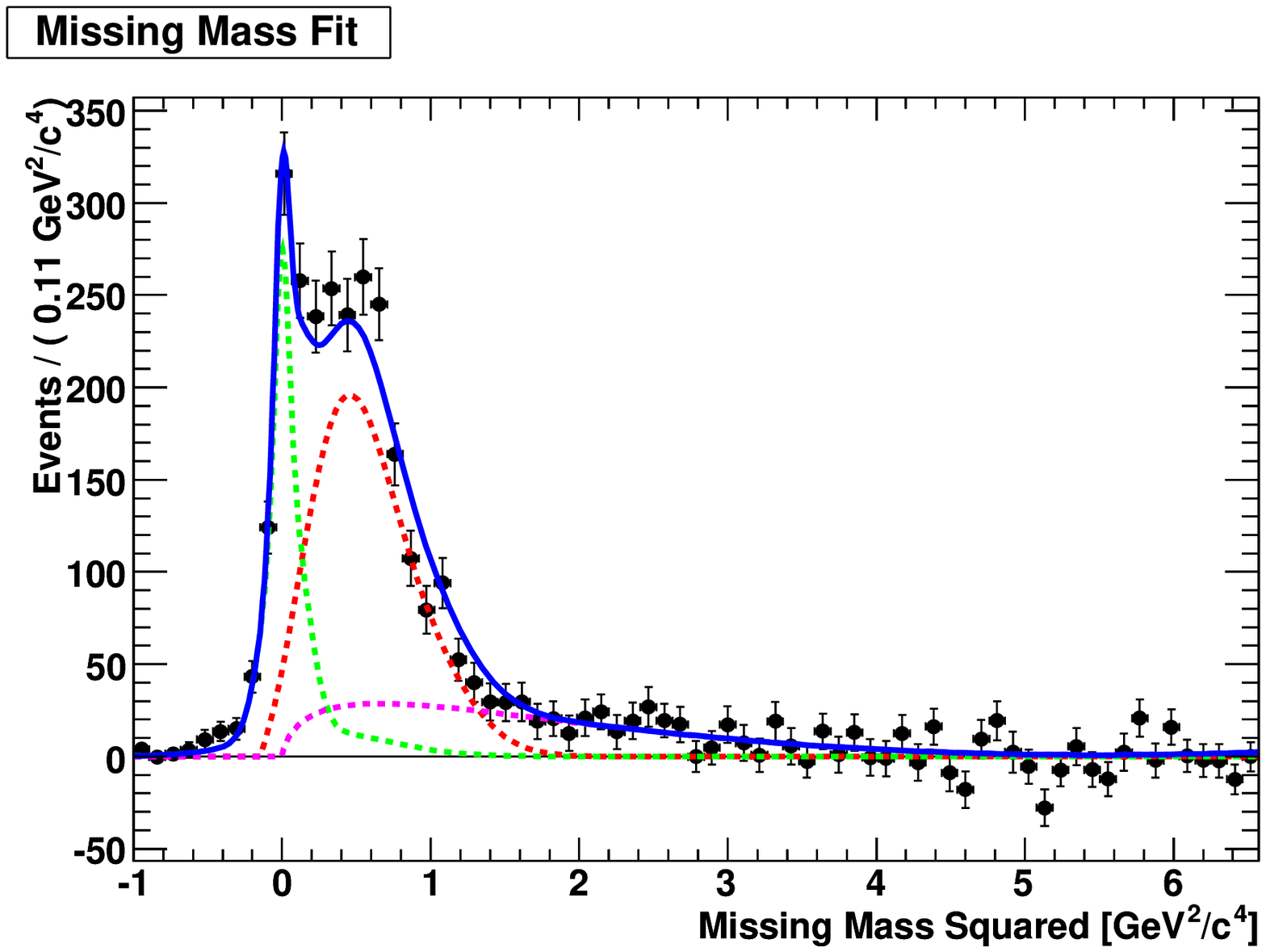,scale=0.4,angle=0}\epsfig{file=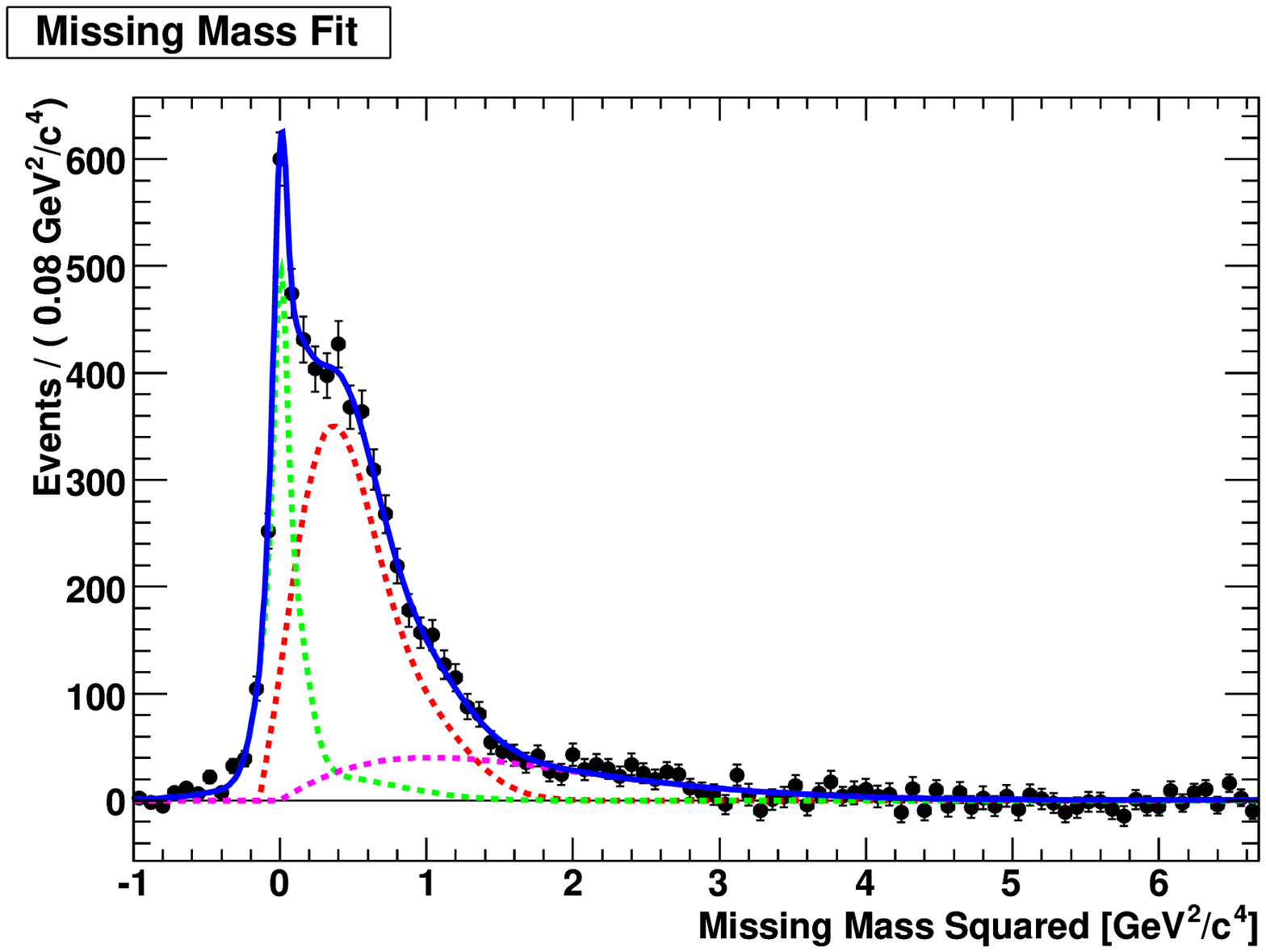,scale=0.4,angle=0}
\caption{
Missing mass  against $Dl$ in full recoil tagged events. Left is $B^0$ decays and right charged $B$ decays.
}
\label{fig:mmdl}
\end{center}
\end{figure}

BABAR~\cite{BABARdl} also makes a measurement of the higher mass states using an inclusive method that also yields estimate
of $Dl\nu$ and $D^*l\nu$.  This is done looking at the semi-leptonic decays recoiling against a fully reconstructed $B$.

Using a recoil tagged sample, the distributions of the missing mass against $Dl$, shown in
Figure~\ref{fig:mmdl}, can be
constructed. The decay modes $Dl\nu,\ D^*l\nu$ and the modes $D^(*)n\pi$ have different shapes in missing mass, 
so the missing mass distribution 
can be used to disentangle them. Roughly speaking, $Dl\nu$ is peaked at zero, $D^*l\nu$ peaks a bit 
higher ($\sim 0.8GeV$) and the rest is broad.

While missing mass is the most powerful variable for discriminating
these decay contributions, there is also discriminating power in the lepton momentum ($p_l$) spectrum 
and in the number of extra tracks not
used reconstructing the $Dl$. 
BABAR extracts the relative branching fractions of these three components by fitting simultaneously to
the missing mass, $p_l$ and extra-track distributions using PDFs determined by the data and validated with the Monte Carlo.
The results are normalized to the semi-leptonic decay rate ($B\rightarrow DXl\nu$). Small contributions from baryons 
in the final state are missing.

The results are summarized in Table~\ref{tab:pegna}. If I use semi-leptonic branching fractions from the 2006 PDG 
(neglecting small non-$DXl\nu$ contributions)
to estimate
absolute branching fractions for $D^*l\nu$, I get $6.34\pm 0.2\%$ for charged $B$'s and $5.58\pm 0.32\%$ for the neutrals.
I've only kept the statistical errors. Under the assumption that most systematics are common, this is again a $\sim 2\sigma$
descrepency...however not all the systematics are common so the discrepancy may not really be that large.

\begin{figure}
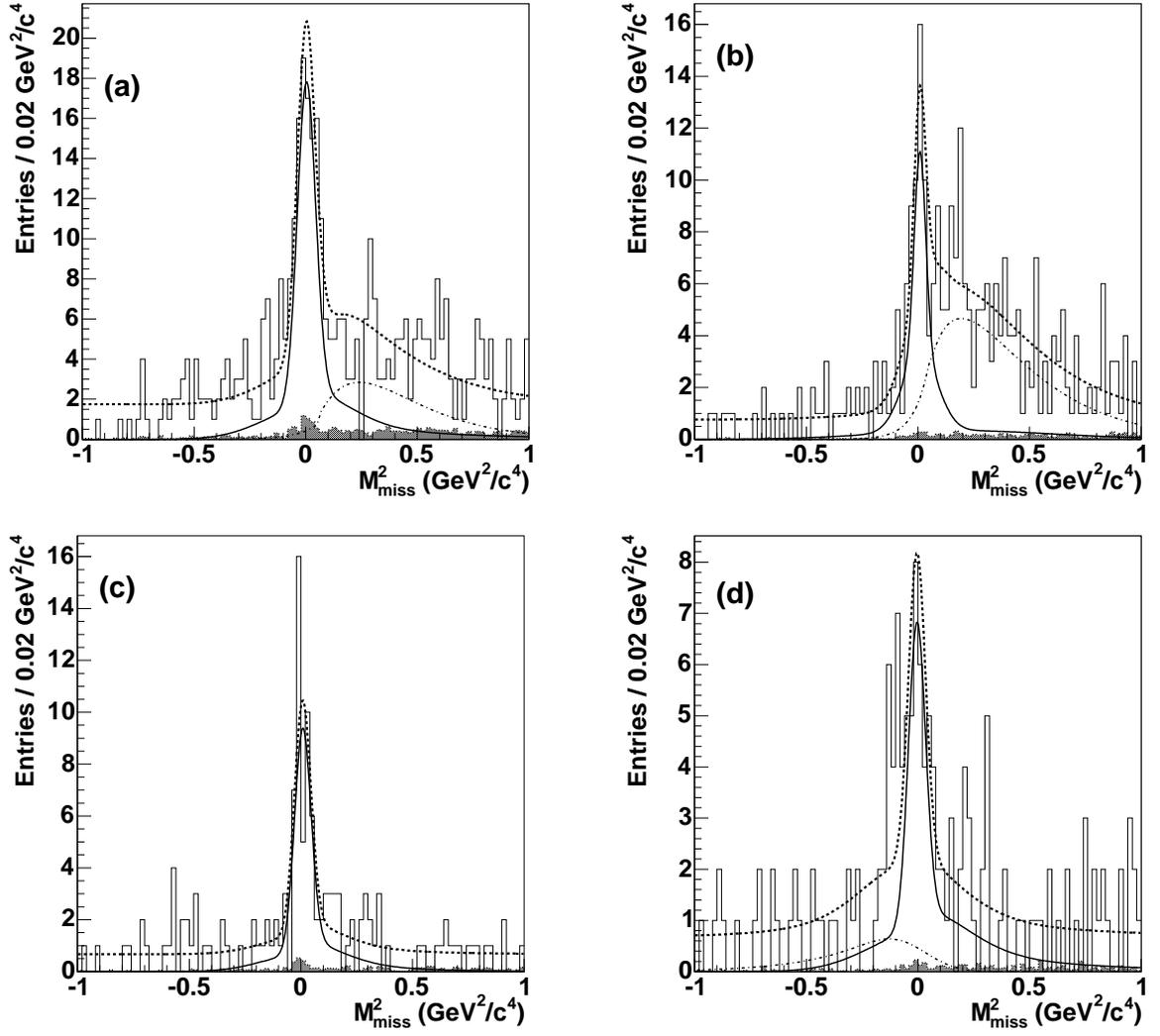

\begin{center}
\epsfig{file=figs/fig3_bc_dpilnu.epsi,scale=0.37,angle=0}\hfill\epsfig{file=figs/fig3_bn_dpilnu.epsi,scale=0.37,angle=0}
\vskip 0.5cm
\epsfig{file=figs/fig3_bc_dstpilnu.epsi,scale=0.37,angle=0}\hfill\epsfig{file=figs/fig3_bn_dstpilnu.epsi,scale=0.37,angle=0}
\caption{
Missing mass  against $D^{*}\pi$ in full recoil tagged events. The modes considered are
(a) $D^+\pi^-l^-\bar{\nu}$, (b) $D^0\pi^+l^-\bar{\nu}$, (c) $D^{*+}\pi^- l^-\bar{\nu}$ and (d) $D^{*0}\pi l^-\bar{\nu}$.
}
\label{fig:mmdpi}
\end{center}
\end{figure}

The BELLE paper~\cite{BELLEA} also contains results on the
higher mass states $D\pi l\nu$ and $D^*\pi l\nu$. In fact, these are the main results of their paper and the $D^* l\nu$
decays are not the focus of the analysis. This technique also employs recoil tagged sample (see Figure~\ref{fig:recoil} again), but
looks at the missing mass recoiling against the specific states. Figure~\ref{fig:mmdpi} show the missing mass 
distribution obtained for the four $D^{(*)}\pi$ modes reconstructed. 
The signal in these plots appears at zero. There is little background in the $D^*\pi$ modes and even in the $D\pi$ modes
the signals are evident.

The branching fractions obtained are
\begin{equation}
B(D^{+}\pi^{-}l^{-}\bar{\nu})=0.54\pm0.07\pm0.07\pm0.06\%,
\end{equation}
\begin{equation}
B(D^{0}\pi^{+}l^{-}\bar{\nu})=0.33\pm0.06\pm0.06\pm0.03\%,
\end{equation}
\begin{equation}
B(D^{*+}\pi^{-}l^{-}\bar{\nu})=0.67\pm0.11\pm0.09\pm0.03\%,
\end{equation}
\begin{equation}
B(D^{*0}\pi^{+}l^{-}\bar{\nu})=0.65\pm0.12\pm0.08\pm0.05\%,
\end{equation}
where in addition to the usual systematic uncertainties there is an additional contribution
from using the $Dl\nu$ and $D^*l\nu$ to normalize the results. The higher mass
contribution to the total branching fraction should be $\sim 10.4-2.1-5.3=3\%$. BELLE's observed $D^{*}\pi$ contribution to
$B^0$ decays is $\approx (0.98\pm 0.13\%)$ which accounts for only $\sim 1/3$. If we assume
\footnote{This is in fact a good assumption, no other isospin is possible given the $c\bar{q}$ produced
must be $I=1/2$.}
 the total isospin of the $D\pi$ system
is $1/2$ then there should be $1/2$ as many again in the unobserved charge $D$ with $\pi^0$ modes for a total, 
of $1.5\pm 0.2$ thus accounting for
$\sim 1/2$ of the missing modes. The same argument applies to the charged $B$, where
the estimate $D\pi$ branching fraction corrected for isospin is $1.8\pm 0.20$. So it seems, like there must
be some contribution from states with two pions. Both of these are consistent with BABAR's estimate using the $Dl$ 
missing mass.

Using isospin to relate $B^0$ and $B^+$ modes we could average to produce a somewhat more accurate estimate. However,
considering the unsettled state of $D^*l\nu$ branching fractions, it's probably best to just settle for the
statement that $D^{(*)}\pi$ can account for $\sim 1/2$ the decays with hadronic masses $m_{had}>m_{D^*}$.

Another interesting number is the ratio $R=D\pi/D^*\pi$. To estimate this I average the BELLE's numbers for $B^0$ and
$B^+$ and assume systematics cancel in the ratio. I find $R=1.58\pm 0.26$ (stat error only).

Using D0's estimate of the branching fraction to the narrow state $D_1^0$ (see above), I find that 
$\sim 35\%$ of $D^*\pi$ are from this narrow state. This suggest that some narrow states might be visible 
in BELLE's recoil samples. 

These are very beautiful measurements. Given that there's still a percent of so in other modes it would be
interesting to measure modes like $D^{*}\pi\pi$.

\begin{table}
\begin{center}
\begin{tabular}{|c|c|c|}
\hline 
Ratio&
$B^{-}(\%)$&
$B^{0}(\%)$\tabularnewline
\hline
\hline 
$Dl\nu$ &
$22.7\pm 1.4\pm 1.5$ &
$21.5\pm 1.6\pm 1.3$
\tabularnewline
\hline 
$D^*l\nu$ &
$58.2\pm 1.8\pm 3.0$&
$53.7\pm 3.1\pm 3.6$
\tabularnewline
\hline 
$D^{*,**}n\pi l\nu$&
$19.1\pm 1.3\pm 1.9$ &
$24.8\pm 3.2\pm 3.0$
\tabularnewline
\hline
\end{tabular}
\label{tab:pegna}
\caption{Ratios of $Dl\nu,\ D^*l\nu$ and $D^{*,**}n\pi$ to $DXl\nu$ for charged and neutral $B$-mesons.}
\end{center}
\end{table}

\section{What to do?}

\begin{figure}
\begin{center}
\epsfig{file=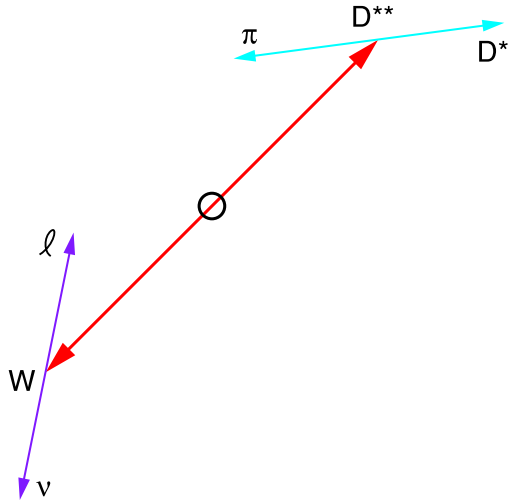,scale=0.45,angle=0}
\caption{
  Possible fake $D^*l\nu$ from a $D^{**}l\nu$ decay -- a neutrino goes ``forward'' and $\pi$ goes ``backward''
the missing mass gets small and the event mimics $D^*l\nu$. Any decay with $\nu$ and $\pi$ moving the same
direction will produce a successful $cos\theta_BY$ construction with just a slightly higher neutrino momentum. The
number of $D^{**}$ decays near this configuration will be influenced by the $D^{**}$ form-factors.
}
\label{fig:fake}
\end{center}
\end{figure}

\begin{itemize} 
\item{We need to resolve the discrepancy between $D^*l\nu$ measured in $B^0$ and $B^+$ decays!}
  \begin{itemize}
  \item{Isospin violation is most unlikely}
  \item{It's only $\sim 2\sigma$, so may be we're just chasing a fluctuation}  
  \item{Could there be high-spin, high-mass states that can mimic $D^{*0}l\nu$ modes?}
    \begin{itemize}
      \item{Try to get solid theory predictions for spin 2 states implemented in MC and
      seek parameters that might cause such a problem, {\it i.e.,} investigate the theoretical constraints on 
      the high mass states.
      }
      \item{Example of ``$D^{**}$'' faking $D^*l\nu$ is shown in Figure~\ref{fig:fake}}
      \end{itemize}
  \item{Maybe slow pions from $D^*\rightarrow D\pi$ decays not well understood?}
    \begin{itemize}
    \item{
      This is not a likely explanation as there aren't many events produced at 
      low $w$ (high $q^2$) so they should not affect branching 
      fraction too much
    }
    \item{Try using $D^+\pi^0$ decay instead of $D^0\pi^+$ to see if the same answer is obtained with a
      different slow $\pi$ efficiency}
    \item{Repeat BELLE style missing mass analysis for $D^*l\nu$ with careful attention to systematics and background.
    Hopefully many systematics would cancel in the charge to neutral $B$ ratios}
    \end{itemize}
  \end{itemize}
\item{$D\pi l\nu$ accounts for about half the missing modes. What is the rest? }
  \begin{itemize}
  \item{Measure $D^{*}\pi\pi$. Is this possible with current statistics?}
  \item{Wide states are hard. They need a full spin-parity analysis to extract phase shifts -- an analysis
  that is not feasible with the current generation of $B$ factories}
  \end{itemize}
\item{Test HQET. Needs higher precision form-factor measurements and ultimately model independent measurements}
    
\end{itemize}

\section{Summary}

We already know quite a bit. The HFAG world average is ${\cal F}(1)_{cb}$ is $36.2\pm 0.6\times 10^{-3}$. 
The form-factor parameters
$R_1$ and $R_2$ have been measured to be $1.42\pm 0.07$ and $0.84\pm 0.043$, respectively. Two narrow contributions
to the high mass region are known at the $10-20\%$ level. It's been established that $D\pi$ (resonant+wide+nonresonant)
can account of $\sim 1/2$ of the branching fraction to higher mass states.

The biggest outstanding problem is {\it tensione} between the $D^*l\nu$ decay in $B^0$ ($\sim 5\%$) and $B^+$ decay
($\sim 6\%$). However, it's only $\sim 2\sigma$, so it probably will resolve itself in due course -- which is not to
say we shouldn't look for a systematic problem in the meanwhile.

While $R_1-R_2$ are well enough measured for current $|V_{cb}|$ measurements 
given the other errors and the theory errors. Improved
measurements could probe the HQET based theoretical assumptions.


\end{document}